# Paradoxical constitutive law between donor O-H bond length and its stretching frequency in water dimer

Rui Liu[1], Xinrui Yang[1], Famin Yu[1] and Zhigang Wang[1,2,3]*

[1]*Institute of Atomic and Molecular Physics, Jilin University, Changchun 130012, China*
[2]*College of Physics, Jilin University, Changchun 130012, China.*
[3]*Institute of Theoretical Chemistry, Jilin University, Changchun 130023, China*
[*]email: wangzg@jlu.edu.cn (Z. W.)

## Abstract

The constitutive laws of hydrogen bonds (H-bonds) are central to understanding microphysical processes not precisely observed, especially in terms of structural properties. Previous experiences in water H-bonding revealed that as the intermolecular O···O distance shortens, the O-H stretching frequency redshifts; thus, an elongated O-H bond length can be empirically inferred, which is described as the constitutive law under the cooperative effect. Here, using the high-precision CCSD(T) method, we report a violation of the conventional constitutive law in water dimer. That is, when the variation in the O···O distance changes from stretching by 0.06 to contracting by -0.15 Å compared to the equilibrium position, the donor O-H bond length decreases from 0.9724 to 0.9717 Å, and the O-H stretching frequency redshifts from 3715 to 3708 cm$^{-1}$. Our work highlights that the O-H bond length decreases simultaneously with its stretching frequency, which is clearly inconsistent with the previously recognized constitutive law.

## Introduction

Constitutive laws are specific relationships that map multiple physical quantities to one another, and for hydrogen-bonding (H-bonding, referring to X-H···Y), more attention is focused on the constitutive law between bond length and vibrational stretching frequency.[1-3] Taking the most prototypical water H-bonding as an example, owing to the limitations of current experimental measurements and characterization techniques, it is difficult to determine the position of the hydrogen atom with sufficient precision,[4] thus direct observation of the changes in O-H bond length below Å, or even 10$^{-2}$ Å, remains a formidable challenge.[5,6] Hence, the O-H bond length can only be measured indirectly through the constitutive law between the O-H bond length and the corresponding stretching frequency.[7,8] This indicates that the constitutive law plays a decisive role in the structural identification responsible for many developments and progress in physics over the last nearly seven decades.[9,10] In fact, this constitutive law has been now empirically well-established that an elongation of the O-H bond length is accompanied by a red-shifted O-H stretching frequency (down-shift),[11] while a shortening of the O-H bond length corresponds to a blue-shifted O-H stretching frequency (up-shift).[12,13] The law was also theoretically depicted using various levels of theory, and the corresponding stretching frequencies were experimentally confirmed.[14-19] Of particular note, this constitutive law is considered as a criterion for defining H-bonds,[11] and is surprisingly robust across a broad variety of H-bonded systems with different H-bond strengths, including water, ion-water and liquid water clusters, etc.[20] Not only that, it has also been employed to determine the long-standing cooperative effect,[21] which is one of the most remarkable features of H-bonds. In water dimer reproducing the structural changes of cooperative effect, this constitutive law was equivalently established that as the O···O distance decreases, the O-H bond length increases, and the concomitant stretching frequency redshifts.[11,22,23] Nevertheless, in 2021, high-precision ab initio calculations on water dimer revealed that as the O···O distance decreases, the O-H bond length always decreases rather than cooperatively increasing, which is referred to as the uncooperative effect of H-bonds.[24] The anomalous uncooperative effect is further attributed to electron correlation. Given the quantum mechanical viewpoint established by the uncooperative effect, the reliability of the constitutive law at the atomic level should be scrutinized seriously.

In this work, we use the well-accepted benchmark method of high-precision ab initio, i.e., coupled-cluster singles and doubles with perturbative triple excitations (CCSD(T)) method, to investigate the constitutive law in the uncooperative water dimer. In particular, detailed analyses showed that when the variation in the O···O distance changes from a stretch of 0.06 to a contraction of -0.15 Å compared to the equilibrium position, a previously-perceived redshift in the O-H stretching frequency corresponds to a decreased O-H bond length arising from the uncooperative effect. That is, the O-H bond length decreases simultaneously with its stretching frequency, which is clearly inconsistent with the previously recognized constitutive law. Our work not only highlights the limitations of the universal constitutive law, but also supports an upper-building exploration associating microphysical processes with previous experimental results.

## Results and discussion

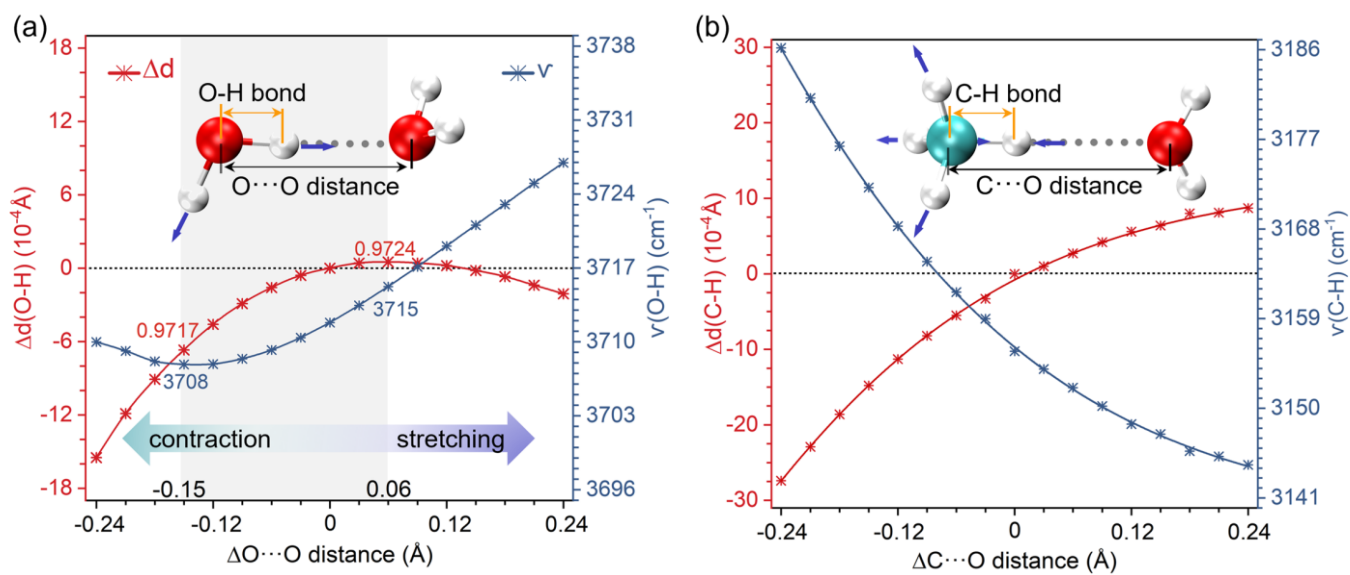

**Figure 1. Geometric structures and intramolecular vibrational stretching frequencies with respect to the O···O (C···O) distance. (a)**, Changes in O-H bond length and symmetric O-H stretching frequencies in HOH···OH$_2$. The grey area highlights the paradoxical constitutive law when the variation in the O···O distance changes from a stretch of 0.06 to a contraction of -0.15 Å compared to the equilibrium position; **(b)**, Changes in the C-H bond length and antisymmetric C-H stretching frequencies in H$_3$CH···OH$_2$. Here, the O···O distance and C···O distance at their relative equilibrium positions are 2.93 and 3.64 Å, respectively. For the variation in the O···O (C···O) distance, negative values indicate contraction, while positive values indicate stretching. For the changes in O-H (C-H) bond length, negative values indicate shortening, while positive values indicate elongation. The equilibrium O-H (C-H) vibrational stretching mode is presented.

We first explored the changes in O-H bond length and respective vibrational stretching frequencies as a function of varying O···O distances for water dimer (see Fig. 1(a)). The O···O distance at the equilibrium position is 2.93 Å, which is taken as the zero point during contraction and stretching. However, the absence of zero-point vibration (ZPV), while not affecting the reliability of the results, brings about stronger H-bonding.[25] This leads to a shortening of the equilibrium O···O distance by approximately 0.04 Å and an up-shift in the equilibrium O-H stretching frequency by approximately 110 cm$^{-1}$ compared to the experimental observations,[6,19,26] which matches well with the theoretical values of previous high-precision ab initio calculations.[27] Here, using the high-precision CCSD(T) method (see Part 1 of the ESI†), we focused on the constitutive law under the uncooperative effect in water dimer. Across the entire plotting range of the variation in the O···O distance (stretching by 0.24 to contracting by -0.24 Å), when the variation in the O···O distance decreases from 0.24 to 0.06 Å, the O-H bond length increases by 3*10$^{-4}$ Å (0.9721 to 0.9724 Å), and the corresponding stretching frequency redshifts (3727 to 3715 cm$^{-1}$). When the variation in the O···O distance decreases from -0.15 to -0.24 Å, the O-H bond length decreases by 9*10$^{-4}$ Å (0.9717 to 0.9708 Å), and the consequent stretching frequency blueshifts (3708 to 3710 cm$^{-1}$). These results are in accordance with the conventional constitutive law between the O-H bond length and the corresponding stretching frequency. However, when the variation in the O···O distance changes from stretching by 0.06 to contracting by -0.15 Å compared to the equilibrium position, we observed a violation of the conventional constitutive law, in which the O-H bond length shortens by 7*10$^{-4}$ Å (0.9724 to 0.9717 Å), due to the recently reported uncooperative effect,[24] the O-H stretching frequency redshifts by 7 cm$^{-1}$ (3715 to 3708 cm$^{-1}$). That is, the O-H bond length decreases simultaneously with its stretching frequency, which is clearly inconsistent with the previously recognized constitutive law. It should be emphasized that the threshold of geometric convergence is further improved to ensure that the accuracy of our calculations is much higher than the O-H bond length increase of 7*10$^{-4}$ Å. Further analysis showed that O-H stretching frequency shift of 7 cm$^{-1}$ is the result of the dominant own mechanism, rather than the vibrational mixing,[28] as displayed in the Part 2 of the ESI†. Since such constitutive law between structural properties and stretching frequency was proposed in the 1950s,[9,10] we found, perhaps for the first time, using CCSD(T) method as benchmark, an unprecedented phenomenon that violates this conventional law. Clearly, the physical mechanism underlying the phenomenon has been attributed to deeper electron correlation in high-precision ab initio calculations, as pointed out before.[24] Therefore, the constitutive law needs to be re-examined.

To further verify the reliability of the violation of the conventional constitutive law, we carried out more test calculations. First, the methane-water dimer was also calculated using a high-precision ab initio method, and the C⋯O distance at the equilibrium position is 3.64 Å. As shown in Fig. 1(b), similar to the water dimer, during the C⋯O distance is also contracted and stretched by the same distance (stretching by 0.24 to contracting by -0.24 Å), the C-H bond length shortens by $3.6*10^{-3}$ Å (1.1030 to 1.0994 Å), and the C-H stretching frequency blueshifts (3144 to 3186 $cm^{-1}$). This result suggests that the methane-water dimer conforms to the conventional constitutive law. It is the violation of the water dimer and the conformity of the methane-water dimer that confirm the reliability of the deeper CCSD(T) calculations. More comprehensive contraction and stretch process are also given in Parts 3 and 4 of the ESI†. It can be found that when the variation in C⋯O distance is further stretched by over 0.45 Å, the weakening of intermolecular interactions leads to a stepwise change of C-H bond from the bound to free, and infrared spectrum appears to be insensitive to the changes of C-H bond length. Clearly, this is probably inevitable for any intermolecular interaction systems.[29] Additionally, we investigated the constitutive law in terms of the hydrogen isotope effect. With the calculations of DOD⋯$OD_2$ and $D_3$CD⋯$OD_2$, the results show that the former also exhibits the analogous paradoxical constitutive law when the variation in the O⋯O distance changes from stretching by 0.06 to contracting by -0.15 Å compared to the equilibrium position (see Part 5 of ESI†). There, the O-D stretching frequency redshifts by 5 $cm^{-1}$ (2685 to 2680 $cm^{-1}$), lower than the 7 $cm^{-1}$ of the O-H stretching frequency in the water dimer, matching well with the theoretical predictions of the hydrogen isotope effect.[30] Importantly, the consideration of the hydrogen isotope effect has no impact on the existence of the paradoxical constitutive law for the water dimer as described above. These results for the water dimer and its hydrogen isotope systems clearly show that the previously-established constitutive law might be problematic and no longer general. In contrast, the latter conforms to the conventional constitutive law, manifesting as the C-D stretching frequency blueshifts by 30 $cm^{-1}$ (2328 to 2358 $cm^{-1}$) when the variation in the C⋯O distance decreases from stretching by 0.24 to contracting by -0.24 Å, which is lower than 42 $cm^{-1}$ of C-H stretching frequency. Thus, for the methane-water dimer, the hydrogen isotope effect does not change the qualitative conclusions.

Overall, we explicitly elucidated the invalidation of the universal constitutive law, at least in the water dimer. The existing theoretical simulations for the constitutive law do not consider complex electron correlations during contraction and stretching based on the high-precision CCSD(T) method.[11-13] Additionally, experimental results can indirectly determine the variation in intermolecular O···O distance accurate to approximately $10^{-2}$ Å by means of matrix isolation, high-resolution vibrational-rotational-tunnelling (VRT), and microwave spectroscopies.[31-33] In contrast, the abovementioned abnormal changes in the O-H bond length resulting from the uncooperative effect are difficult to obtain through the current experimental precision, which also implies an unknown reason for the paradoxical constitutive law under the uncooperative effect. Furthermore, given that both the O···O (C···O) distance and the corresponding stretching frequency are expected to be observed experimentally, the reliability of the constitutive law between these two components should also be profoundly validated in the following.

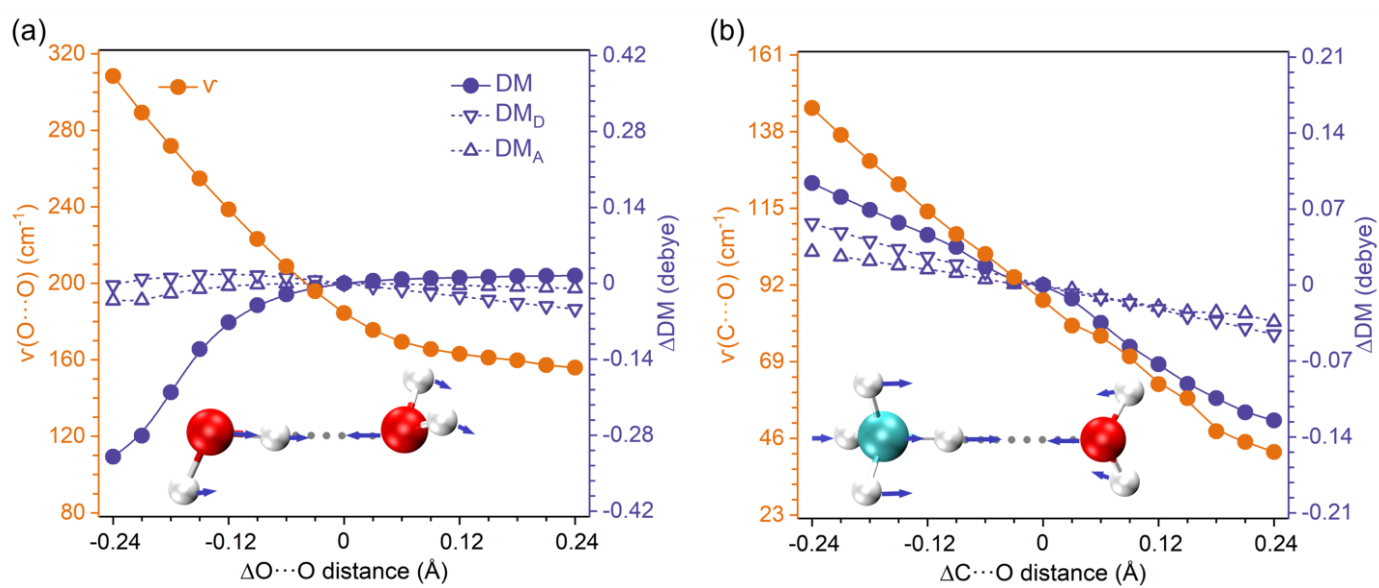

**Figure 2. Intermolecular vibrational stretching frequencies and dipole moments with respect to the O···O (C···O) distance. (a)**, Changes in the O···O stretching frequencies and dipole moments for the water dimer and two water monomers in HOH···OH$_2$. **(b)**, Changes in C···O stretching frequencies and dipole moments for methane-water dimer and methane and water monomers in H$_3$CH···OH$_2$. The equilibrium O···O and C···O vibrational stretching modes are presented.

Based on the above, the constitutive law between the intermolecular O⋯O distance and its stretching frequency in the water dimer was further discussed. As shown in Fig. 2(a), over the entire plotting range of the variation of O⋯O distance changing from stretching by 0.24 to contracting by -0.24 Å, the O⋯O stretching frequency blueshifts (156 to 308 cm$^{-1}$). Furthermore, the methane-water dimer was also studied, showing a blueshift in the C⋯O vibrational stretching frequency (42 to 145 cm$^{-1}$) when the variation in the C⋯O distance changes from stretching by 0.24 to contracting by -0.24 Å (see Part 6 of ESI†). These results indicate that the intermolecular O⋯O (C⋯O) distance decreases and the corresponding O⋯O (C⋯O) stretching frequency increases, implying that the change in the O⋯O (C⋯O) distance can be further experimentally determined by the O⋯O (C⋯O) stretching frequency. It is verified that the constitutive law between the intermolecular O⋯O (C⋯O) distance and corresponding stretching frequency is not broken. In this way, shifting the previously greater focus on the structural properties and stretching frequencies of intramolecular O-H (C-H) bond in H-bonds to a focus on those of intermolecular O⋯O (C⋯O), etc., which would be promising for developing a more reliable constitutive law. Of course, a non-negligible problem arises that, their frequency ranges are very different.

Next, considering that the dipole moment, as an intrinsic molecular property, can be obtained by measuring the Stark effect, and that in the water dimer, the dipole moment has also important implications for structural determination,33-35 this parameter can provide another form of comparison between theory and experiment. Thus, we verified the constitutive law again from the perspective of the dipole moment, as depicted in the Fig. 2(b). For simplicity, the dipole moments of dimer, donor and acceptor monomers are labelled as DM, DMD and DMA, respectively. Specifically, as the variation in the O⋯O distance decreases from stretching by 0.24 to contracting by -0.24 Å, the DM of the water dimer decreases intuitively (2.69 to 2.35 D). The DMD and DMA initially increases (2.11 to 2.12 D and 2.22 to 2.28 D) and then decreases (2.12 to 2.08 D and 2.28 to 2.26 D) when the variation in the O⋯O distance is contracted by over -0.03 Å and contracted by over -0.12 Å, respectively. Importantly, the dipole moment also fails to response to the anomalous O-H bond length change. For the methane-water dimer, as the variation in the C⋯O distance decreases from stretching by 0.24 to contracting by -0.24 Å, the DM of the methane-water dimer increases sustainably (2.16 to 2.38 D) and the DMD and DMA of methane and water monomers also increase (0.16 to 0.26 D and 2.06 to 2.12 D), as shown in Part 7 of the ESI†. It can be seen that the dipole moments of both water and methane-water dimers change monotonously as the intermolecular O···O (C···O) distance decreases, and this trend is also reflected in previous theoretical studies on other monomers with C-H bonds and water clusters.[29,36] Therefore, it remains feasible and instructive to characterize the inter- and intramolecular structural properties by the dipole moment.

## Summary

In conclusion, we investigated the paradoxical constitutive law between the donor O-H bond length and its stretching frequency in water dimer. Our high-precision ab initio calculations showed that when the variation in O···O distance changes from stretching by 0.06 to contracting by -0.15 Å relative to the equilibrium position, a decreased O-H bond length (0.9724 to 0.9717 Å) resulting from the uncooperative effect dramatically yields a previously-measured red-shifted O-H stretching frequency (3715 to 3708 cm$^{-1}$). Obviously, the O-H bond length decreases simultaneously with its stretching frequency, which violates the conventional constitutive law followed by the uncooperative methane-water dimer. For comparison, we also calculated the O···O stretching frequency and dipole moment. Although these results are consistent with the conventional constitutive law, they do not alter our principal conclusions about the changes in uncooperative O-H bonds in water dimer. Our work therefore suggests that more precise experimental tools are urgently required to deeply understand and characterize complex intermolecular interactions, especially H-bonding, enabling a reliable constitutive law to be obtained at the atomic level.

## Methods

In this work, all structures were optimized at the CCSD(T) level[37] with the aug-cc-pVDZ basis set. The CCSD(T) method has the advantage of computational efficiency and accurately explores the structures and properties of H-bonds; thus, it is widely employed for the calculations of H-bonding. To ensure the reliability of our conclusions, the threshold for SCF convergence of 10$^{-10}$ a.u. was used for all of the calculations. And the maximum component gradient was set to less than 10$^{-5}$ a.u. for optimizing different variables. The geometric structures were relaxed at each given intermolecular distance during contraction and stretching of the distance in steps of 0.03 Å (constrained in optimization). The vibrational frequency, using the numerical Hessian energy gradient, was calculated at the same level. Further analysis by setting different displacements also confirms the reliability of the calculated frequency. In addition, the dipole moments in different structures were analysed at the same level. Above all calculations were performed using the MOLPRO 2012 program.[38] Moreover, the vibrational mixing of O-H stretching frequency is analyzed based on partial vibrational spectrum (PVS) analysis using Multiwfn program[39] in HOH⋯OH$_2$.

## Author contributions

R. Liu performed most of simulations. Z. Wang. conceived this project. R. Liu, X. Yang, F. Yu and Z. Wang analyzed the results. R. Liu and Z. Wang contributed to writing the paper. All co-authors discussed the results and commented on the manuscript.

## Competing interests

All authors declare no competing interests.